# $t \to c\gamma$ decay in the ACD model

G. G. Devidze, A. G. Liparteliani

High Energy Physics Institute, Tbilisi State University,
University St.9, Tbilisi 0186, Georgia
email adresses: gela.devidze@tsu.ge, a.liparteliani@hepi.edu.ge

*Abstract:*

In the presented paper we have estimated additional to the standard model (SM) contributions to the top quark FCNC decays in the framework of the model with one extra dimension. The estimates of performed calculations show that extra dimension models with only one additional spatial dimension cannot raise branching ratios for top FCNC decays.

Success of the Standard Model (SM) do not weaken theoretical arguments in favour of New Physics (NP), which is aticipated at the TeV scale. The task to find and identify NP seems to us to be as a most important challenge for High Energy Physics. In spite of successes of SM of particle physics and SM of cosmology (based on traditional theory of General Relativity), there are profound experimental and theoretical reasons to suppose that both of them are incomplete. From the experimental point of view to this opinion hint, for example, the small but non vanishing neutrino masses and small but non vanishing value for cosmological constant, the presence of dark matter, dark energy and baryon assimmetry of the Universe. Theoretical problems include hierarchy problem, supersimmetry breaking, replication of fermion generations and highly hierarchical structure of fermion mass matrixes, $CP$ − conservation in QCD etc.

Very important is the quuestion on how to identify the manifestation of NP and how to give the preference to some special kind of SM expansion. As a most promising extentions of the SM are considered SUSY extentions, based on the idea of supersimmetry [1,2,3,4] and approaches with large extra dimensions [5-9], though there are also more radical ways beyond SM.

In the framework of extra dimensional models with the fundamental gravity scale around ~TeV the NP is expected to manifest itself around this scale. As soon as indicated scale will be tested, for example at LHC energies, NP must manifest itself. The question is only in which form it manifests itself. The most direct way to manifest and analyze experimental patterns of NP could consist in the direct production of the new particles like supersimmetric or Kaluza-Klein (KK) resonances. Another possibility consists in the indirect manifestation of the effects beyond SM. Before showing in the direct production processes, beyond SM effects could manifest themselves in the rare decays through the various loop effects. The importance of such a possibility is hard to be overestimated. To differ various NP scenarios, we are in need to investigate their influence on the aromatodynamics.

FCNC processes are strongly suppressed in the SM. So, there are even no theoretical hint for FCNC in the lepton sector in frame of the SM. In spite of the success of the SM of particle physics there are profound reasons to think that it is incomplete. On the experimental side, these are very small but non-vanishing neutrino masses, small but non-zero cosmological constant, the presence of dark matter, dark energy and the baryon asymmetry of the universe[10]. In the quark sector FCNC processes first arize on the one loop level and are suppressed by GIM mechanism. Some, of such processes ($B^0 \to \overline{B}^0, K^0 \to \overline{K}^0, B \to K^*\gamma$) are experimentally observed and are in quite good agreement with SM predictions, though the last experimental data on $B^0 \to \mu^+\mu^-$ decays could hint on the NP effect. As for top quark sector, the FCNC processes in it are also very small in the SM with following rates[10]:

$$Br(t \to cg) \approx 5 \cdot 10^{-11}, \quad Br(t \to c\gamma) \approx 5 \cdot 10^{-13}, \quad Br(t \to cZ) \approx 1.310^{-13}, \quad Br(t \to cH) \approx 10^{-13}$$

So, even single experimental observation for any top rare processes could be breakthrough beyond the SM physics. In some extensions of the SM there could be several mechanisms for enhancements of FCNC processes, including top quark FCNC phenomena. One of such mechanism could be arisen in case when GIM suppression changes its quadratic nature by linear one. This happens when particles running in the loops have comparable masses. Such a situation is realized in the UED models.

The modern models [5-8] with extra space-time dimensions have received a great deal of attention because the scale at which extra dimensional effects can be relevant could be around a few TeV. The first proposal for using large (TeV) extra dimensions in the SM with gauge fields in the bulk and a matter localized on the orbifold fixed points was developed in Ref. [9]. Models with extra space-time dimensions can be constructed in several ways. Among them the following major approaches are most remarkable: i) the ADD model of Arkani-Hamed, Dimopoulos and Dvali [5,6]; in this approach all elementary particles except the graviton are localized on the brane, while the graviton propagates in the whole bulk; ii) the RS model of Randal and Sundrum with a warped 5-dimensional space-time and nonfactorized geometry [7]; iii) the ACD model of Appelquist, Cheng and Dobrescu (so called Universal Extra Dimensional model), where all the particles move in the whole bulk [8].

In the Universal Extra Dimension (UED) scenarios all the fields presented in the SM live in extra dimensions, i.e. they are functions of all space-time coordinates. For bosonic fields one simply replaces all derivatives and fields in the SM lagrangian by their five dimensional counter parts. These are the $U(1)_Y$- and $SU(2)_L$- gauge fields, as well as the $SU(3)_C$-gauge fields from the QCD sector. The Higgs doublet is chosen to be even under $P_5$ ($P_5$ is a parity operator in the five dimensional space) and possesses a zero mode. Note that all zero modes remain massless before the Higgs mechanism is applied. In addition we should note that as a result of the action of the parity operator the fields receive additional masses ~n/R after dimensional reduction and transition to the four dimensional Lagrangians.

Remarkable feature of the universal type models of extra dimension is the preservation of so called KK- parity, which results in the absence of Kaluza-Klein contributions at the tree level into processes, which take the place at scales $\mu << 1/r$, where $r$ is the compactification radii for the extra dimension. In a certain sense the KK-parity conservation is like to preservation of R-parity in the supersimmetryc theories.

From the analysis of the precision electroweak observables we could conclude that $1/R \geq 700-800$ Gev. Investigations of the decay $b \to s\gamma$ demonstrate the result that $1/R \geq 250$ GeV, at least. Moreover, from the investigation of $B \to K^*\gamma$ we have $1/R \geq 250$ GeV, while inclusive decays $\bar{B} \to X_s\gamma$ show that $1/R \geq 600$ GeV. Several UED contributions have been studied in various FCNC processes. These investigations indicate that $1/R \geq 300$ GeV. Finally, we can conclude that lowest upper limit for $1/R$ could be 250 GeV.

Namely, the conservation of KK-parity prohibits single production of zero KK-modes in the processes of interaction for "usual" particles. From this point of view the flavor change neutral current (FCNC) inspired processes like ($K^0 - \bar{K}^0$, $B^0 - \bar{B}^0$), rare B– and K–decays or radiative decays in the quark or leptonic sectors are most interesting because they are kind of processes which, not having their tree level analogues, first arise in the SM at one loop level, and hence are strongly suppressed.

In the five dimensional ACD model the same procedure for gauge fixing is possible as in the models in which fermions are localized on the four dimensional subspace. With the gauge fixed one can diagonalize the kinetic terms of the bosons and finally derive expressions for the propagators. Compared to the SM, there are additional Kaluza-Klein (KK) mass terms. As they are common to all fields, their contributions to the mass matrix of gauge bosons are proportional to the unity matrix. As a consequence, the electroweak angle remains the same for all KK modes and is the ordinary Weinberg angle $\theta_W$. Because of the KK-contribution to the mass matrix, charged and neutral Higgs components with $n \neq 0$ (n being the number of the KK-mode) no longer play the role of Goldstone bosons . Instead, they mix with $W_5^\pm$ and $Z_5$ to form, in addition to the goldstone modes $G_{(n)}^o$ and $G_{(n)}^\pm$, three additional physical states $a_{(n)}^o$ and $a_{(n)}^\pm$. The charged particles $a^{\pm}_{(n)}$, $G^{\pm}_{(n)}$ and the towers of W-boson contributes into top FCNC decays in the ACD model (Fig.1).

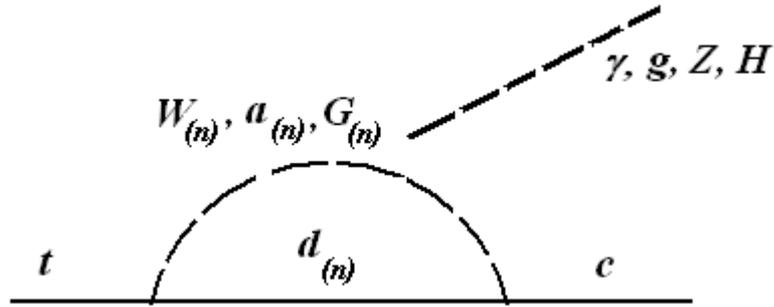

Fig.1. One-photon radiative decay $t \to c\gamma$ in the ACD model via intermediate particles $a^{\pm}_{(n)}$, $G^{\pm}_{(n)}$, $W_{(n)}$.

FV processes are intensively investigated in large extra dimension scenarios. As these studies show in case when theoretical approaches are not enriched other way than simply adding extra dimension to the SM, there is hard to get theoretical predictions close to experimental bounds [11-16].

It is expected from common theoretical sense that FV processes would be possibly enhanced in case when particles running in the appropriate loops have close masses [17]. Loop amplitudes with comparable masses of inter mediate particles running in the loop seem to be quite large because the generic quadratic suppression factor is changed to a linear one. Such a situation with comparable masses in principle is realizable in the models with extra dimensions. It is not obvious without specific calculations how would be changed the SM estimate of the above processes in the models with extra dimensions. Some details of the models can enhance suitable amplitudes and others can cause suppression. On general grounds, one expects an enhancement of the amplitudes, but this expectation is not fulfilled because of the almost degeneracy of the massive towers modes from different generations. This is not necessarily the last word, though; the black hole can inspire FV processes and enhance them [18-22]. On the other hand it is not obvious without specific calculations how would be changed the SM estimate of the above processes in the models with extra dimensions (ED). Some details of the models can enhance suitable amplitudes and others can cause suppression. It is impossible to estimate summary effects of this interplay without specific calculations. We have calculated the relevant contributions to t→cγ via intermediate black hole.

We accept the conjecture that black holes violate global symmetries [21-22] including lepton family number. So, black holes could manifest themselves in LFV processes as intermediate states and enhance them. We assume that black holes with mass lighter than effective Planck mass have a zero charges (electric, color) and zero angular moment in the classical case and this feature is adopted by quantum gravity too. So, one can write the effective Lagrangian describing interactions between quarks and black hole in the following way [21, 22]

$$L = g_{ij}\bar{q}_{Li}q_{Rj}\Phi_{bh} + h.c. \qquad (1)$$

where $g_{ij}$ are dimensionless effective coupling constants. Virtual black hole can induce FV process $t \to c\gamma$ at the one loop level (Fig.2).

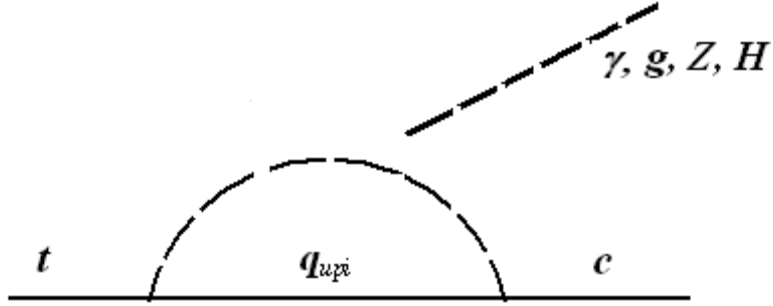

FIG. 2. One-photon radiative decay $t \to c\gamma$ via intermediate black hole (dashed intermediate line).

Through direct calculation we get following expression for the decay width

$$\Gamma(t \to c\gamma) = \frac{\alpha m_t^5}{(4\pi)^4 M^4}\left(g_{ei}g_{\mu i}^* f(x_i)\right)^2 \qquad (2)$$

where $M$ is mass of the intermediate black hole and we have introduced following notation

$$f(x) = \frac{1}{36(1-x)^4}(2+3x-6x^2+x^3+6x\ln x), \quad x_i = \frac{m^2(q_{up_i})}{M^2} \qquad (3)$$

For the total decay width of the top quark we have used $\Gamma = 1.55 GeV$. Using formulae (2), (3) one get

$$Br(t \to c\gamma) = 4.4\cdot 10^{-8}\left(g_{ti}g_{ci}^* f(x_i)\right)^2 \qquad (4)$$

As numerical analyses show the branching ratio does not increases.

**Conclusion**

Experimental success of SM is very expressive in the last decades after its acknolwledgment. At least we know only experimental derivation from "standard thinking" due to discovery of small finite (but nonzero) neutrino masses in the various neutrino oscillation experiments. Therefore it is important to know how numerous and at which confidence level will be realized intervention of any kind of new physics (NP) beyond SM in all sectors of the knowledge of high energy physics. We have discussed one of the ways on the theoretical road for manifestation of NP in the top quark sector, particularly, in the top FCNC decays. As our analyses show the ED models with only one additional spatial dimension cannot raise branching ratios for top FCNC decays. Experimental observation of a few events with these decays opens the possibility for NP intervention different from the ED.

**REFERENCES**

1. Yu. A. Golfand and E. P. Likhtman, JETP Lett. 13(8) (1971) 452,

2. D. V. Volkov, V. P. Akulov, JETP Lett. 16(11) (1972) 621,

3. D. V. Volkov, V. P. Akulov and V. A. Soroka, JETP Lett. 22(7) (1975) 396,

4. J. Wess and B. Zumino, Nucl. Phys. B **70**, 39 (1974).

5. N. Arkani-Hamed, S. Dimopoulos, and G. Dvali, Phys.Lett. B **429**, 263 (1998),

6. I. Antoniadis, N. Arkani-Hamed, S. Dimopoulos, and G. Dvali, Phys. Lett. B **436**, 257 (1998),

7. L. Randall and R. Sundrum, Phys. Lett. **83**, 3370 (1999),
   L. Randall and R. Sundrum, Phys. Lett. 83, 46090 (1999),

8. T. Appelquist, H.C. Cheng, and B.A. Dobrescu, Phys. Rev. D **64**, 035002 (2001).

9 .I.Antoniadis, Phys. Lett. B46 (1990) 317.

10. Particle Data Group, http://pdg.lbl.gov.

11. G.A. Gonzalez-Sprinberg, R. Martinez, Jairo Alexis Rodriguez, Eur. Phys. J. C51 (2007) 919-926.

12. G. Devidze, A. Liparteliani, GESJ Phys. 2012 N1 (2012) 28-33.
13. G.Devidze, GESJ Phys. 2009 N1 (2009) 13-21
14. G.G. Devidze, A. G. Liparteliani, V.G. Kartvelishvili, GESJ Phys. 2010N2 (2010) 9-14

15. G.G. Devidze, A. G. Liparteliani, Phys.Atom.Nucl. 70 (2007) 1131-1135.

16. G.G. Devidze, A. G. Liparteliani, Phys.Atom.Nucl. 69 (2006) 1538-1541.

17. Bo He, T.P. Cheng and Ling-Fong Li, Phys. Lett. **B553**, 277 (2003).

18. S. Dimopoulos and G. Landsberg, Phys. Rev. Lett. 87, 161602 (2001).

19. S.B. Giddings and S.D. Thomas, Phys. Rev. D 65, 056010 (2002).

20. D. Stojkovic, G.D. Strakman, and F.C. Adams, Int. J Mod. Phys. D 14, 2293 (2005).

21. C. Bambi, A.D. Dolgov, K. Freese, Nucl. Phys. B 763, 91 (2007).

22. J. Doukas, S.R. Choudhury and G.G. Joshi, Mod. Phys. Lett. A 21, 2561 (2006).